\documentclass{tPHM2e}

\usepackage{graphicx}
\usepackage{bm}
\usepackage{amsmath}
\usepackage{amssymb}

\newcommand{\be}[1]{\begin{equation}\label{eq:#1}}
\newcommand{\ee}{\end{equation}}
\newcommand{\bea}{\begin{eqnarray}}
\newcommand{\eea}{\end{eqnarray}}

\newcommand{\phd}{\phantom{\dag}}
\newcommand{\ph}{\phantom{.}}
\newcommand{\up}{^{\phd}}
\newcommand{\noi}{\noindent}
\newcommand{\no}{\nonumber}

\usepackage{color}

\begin{document}
\doi{10.1080/14786435.20xx.xxxxxx}
\issn{1478-6443}
\issnp{1478-6435}
\jvol{00} \jnum{00} \jyear{2010} 



\title{Magnetic-field-induced chiral hidden order in $\bm{{\rm URu_2Si_2}}$}

\author{P. Kotetes$^{\rm a}$ $^{\ast}$ \thanks{$^\ast$Corresponding author. Email: panagiotis.kotetes@kit.edu
\vspace{6pt}}, A. Aperis$^{\rm b}$ and G. Varelogiannis$^{\rm c}$\\\vspace{6pt} 
$^{\rm a}${Institut f\"{u}r Theoretische Festk\"{o}rperphysik and DFG-Center for Functional Nanostructures (CFN), Karlsruhe Institute of Technology, 76128 Karlsruhe,
Germany}\\
$^{\rm b}${Department of Physics and Astronomy, Uppsala University, Box 516, SE-75120 Uppsala, Sweden}\\
$^{\rm c}${Department of Physics, National Technical University of Athens, GR-15780 Athens, Greece}
\\\vspace{6pt}}

\maketitle

\begin{abstract}
Two of the most striking and yet unresolved manifestations of the hidden order (HO) in ${\rm URu_2Si_2}$, are associated on one hand with the double-step metamagnetic
transitions and on the other with the giant anomalous Nernst signal. Both are observed when a magnetic field is applied along the c-axis. Here we provide for the first time a
unified understanding of these puzzling phenomena and the related field-temperature (${\cal B}-T$) phase diagram. We demonstrate that the HO phase at finite fields can be
explained with a chiral $d_{xy}+id_{x^2-y^2}$ spin density wave, assuming that the zero field HO contains only the time-reversal symmetry preserving $id_{x^2-y^2}$
component. We argue that the presence of the field-induced chiral HO can be reflected in a distinctive non-linear ${\cal B}$-dependence of the Kerr angle, when a Kerr
experiment is conducted for finite fields. This fingerprint can be conclusive for the possible emergence of chirality in the HO.
\bigskip

\bigskip

\end{abstract}

\section{Introduction}

The `hidden' order (HO) transition of heavy fermion ${\rm URu_2Si_2}$ \cite{MydoshRev} attracted enormous attention due to the mismatch of its specific heat jump 
at T$_{\rm o}$=17.5K and the concomitant c-axis oriented tiny antiferromagnetic moment $0.03\mu_B/U$, of wave-vector $\bm{Q}_0=(0,0,1)$ \cite{Palstra,Schlabitz}. This
incompati\-bi\-li\-ty has raised great controversy on whether the genuine spin-dipolar antiferromagnetic moment inferred from the Bragg reflection at $\bm{Q}_0$ \cite{Walker},
may be the driving order parameter of this transition. In fact, NMR measurements \cite{AmitsukaInhomogeneousMagnetism,NoSMAF} and Larmor diffraction \cite{Niklowitz}, suggest
that antiferromagnetism is inhomogeneous and parasitic in the HO phase. In addition, experiments have found no evidence for higher spin-correlators \cite{Mason}.

Inelastic neutron scattering (INS) measurements \cite{Broholm,Bourdarot,WiebeExcitations,Aoki,AokiReentrantHO}, have revealed two gapped spin excitations at the wave-vectors
$\bm{Q}=(1,0,0)$ (equivalent to $\bm{Q}_0$) and $\bm{Q}_1=(1\pm 0.4,0,0)$. Specifically, the first resonance always accompanies the HO phase and disappears in the
pressure-induced large moment antiferromagnetic (LMAF) phase \cite{Aoki,AokiReentrantHO} of ${\rm URu_2Si_2}$. In stark contrast, the second persists and acquires a
larger gap upon effecting pressure. Moreover, the emergence of incommensurate excitations can be only understood within an iti\-ne\-rant picture \cite{WiebeExcitations},
clearly contradicting the predictions of a number of theo\-retical models, solely based on the localized nature of the U-5f electrons
\cite{Santini,Gorkov,Kiss,ArrestedKondo,hastatic}. The itinerant perspective
\cite{Ramirez,Ikeda,Chandra,Virosztek,Oppeneer,Varma,IncommensurateSpinResonance,Triakontadipole,Oppeneer3,FujimotoPRL,Das,Riseborough,IkedaR5} is also supported by the
existence of strong Fermi surface ne\-sting at $\bm{Q}_1\up$ \cite{Oppeneer,Oppeneer2} and the recent verification of Fermi surface gapping in Angularly Resolved Photoemission
Spectroscopy (ARPES) experiments \cite{arpes,OppeneerARPES}. 

Under strong external magnetic fields applied along the c-axis and at sufficiently low temperature, ${\rm URu_2Si_2}$ exhibits a cascade of transitions from the HO to a  
multitude of unidentified phases that bear similarities with the HO \cite{Kim, reentrantHO}. The sharpest of these transitions are from the HO to a re-entrant HO (RHO) phase
around 35T and to a polarized paramagnetic phase near 39T. These are accompanied by double-step itinerant metamagnetism, i.e. two first order jumps in the magnetization;
one entering and another one leaving the RHO regime \cite{reentrantHO}. At e\-le\-va\-ted temperatures, the transitions become second order, signalling the presence of a
metamagnetic critical end point (MCEP) in the ${\cal B}-T$ phase diagram of this material. This intricate behavior, is reminiscent to the case of ${\rm Sr_3Ru_2O_7}$
\cite{Grigera}, a prototypical itinerant metamagnet, thus providing further support for the importance of itinerant electron physics in the HO phase.

Apart from the remarkable changes in the phase diagram of the HO, transport measurements under applied magnetic fields inside the HO have revealed that this compound is 
characterized by an anomalous thermoelectric behaviour up to 35T \cite{Bel,Behnia}, where the HO has been considered to collapse \cite{Behnia}. The arising Nernst signal has a
po\-si\-tive sign and a giant magnitude of several ${\rm \mu V/K}$, while the temperature evolution of the Nernst response resembles to a `tilted-hill', demonstrating a peak
at about 3-4K, where it reaches the value of $30{\rm \mu V/K}$ for ${\cal B}\simeq 12$T. Up to now, the enhanced Nernst signal has been solely viewed as the signature  of the
semi-metallic character of this electron-hole compensated metal \cite{Bel,Behnia}.

Here, we propose that the HO in the ${\cal B}-T$ phase diagram of ${\rm URu_2Si_2}$ is a chiral $d_{xy}+id_{x^2-y^2}$ spin density wave (d-SDW) phase. Within a
mean-field approximation we reproduce self-consistently the experimentally resolved ${\cal B}-T$ phase diagram of this compound. Moreover, we provide a unified explanation for
the emergence of cha\-ra\-cte\-ri\-stic phenomena that accompany the HO phase such as the puzzling metamagnetism \cite{reentrantHO,Kim} and the giant Nernst signal
\cite{Bel,Behnia}. In addition, we propose a Kerr angle measurement in the presence of an external magnetic field, that could directly identify the formation of a chiral
HO in this compound, through the \textit{non-linear} ${\cal B}$-dependence associated with the induced chirality. Note that our proposal is also compatible with the
chiral supercon\-duc\-ting state that is believed to appear within the HO \cite{Sigrist}.

\section{Mean-field description of the field induced chiral d-SDW HO}

The particular $d_{xy}+id_{x^2-y^2}$ momentum structure of the order parameter is consi\-de\-red here to originate from repulsive nea\-rest and next-nea\-rest inter-site
Coulomb interactions of the form $\sum_{\bm{i}\neq\bm{j}}V_{\bm{i}\bm{j}}\up n_{\bm{i}}\up n_{\bm{j}}\up$, while it is described by a two component c-axis polarized
spin-triplet order parameter $\Delta^z(\bm{k})=\Delta_1\sin k_x\sin k_y-i\Delta_2(\cos k_x-\cos k_y)$, with the second of them bearing similarities to orbital orders already
proposed for URu$_2$Si$_2$ \cite{Ramirez,Ikeda,Chandra,Virosztek,FujimotoPRL,Oppeneer3}. The additional $d_{xy}$ component renders the HO \textit{chiral} rather than
\textit{orbital}, vio\-lating time-reversal symmetry in a macroscopic level due to the development of an intrinsic orbital ferromagnetic moment. It has been shown, that due to
this orbital magnetic-field-coupling, any unconventional density wave of the form $id_{x^2-y^2}$ will acquire an additional field-dependent $d_{xy}$ component
\cite{Balatsky,FICDDW}. Even if the chiral order does not appear in zero-field, it will be necessarily field-induced, affecting the ${\cal B}-T$ phase diagram of the HO and
e\-ve\-ry transport measurement realized in the presence of a field, such as the Nernst effect. In fact, when one of the two components is solely field-induced, the chiral
order may be viewed as a magnetic-field-renormalized orbital order (see Appendix). In this sense, our theory is fully compa\-ti\-ble with all the earlier proposed
orbital-order-based models \cite{Ramirez,Ikeda,Chandra,Virosztek,FujimotoPRL,Oppeneer3} and it could be considered as a natural and necessary extension to them. 

However, the chiral order exhibits properties not familiar to orbital orders. Ge\-ne\-ral\-ly the spontaneous appearance of a chiral phase leads to a polar Kerr effect
\cite{PolarChiral} and the ge\-ne\-ra\-tion of anomalous Hall electric \cite{SQHE} and thermoelectric currents \cite{ANEYakovenko,ChiralityNernst}, inclu\-ding the
`tilted-hill' giant Nernst signal recently shown to arise in strongly insulating chiral condensates \cite{ChiralityNernst}. In the present case, the chiral HO will be
field-induced and it will manifest its presence in all the phenomena mentioned above through the characteristic non-linear field-dependence of the $d_{xy}$ order parameter.

\section{Metamagnetism within our self-consistently extracted magnetic field-temperature phase diagram}

\begin{figure}[h]\centering
\begin{minipage}[b]{4.8in}
\includegraphics[width=0.45\textwidth,height=0.34\textwidth]{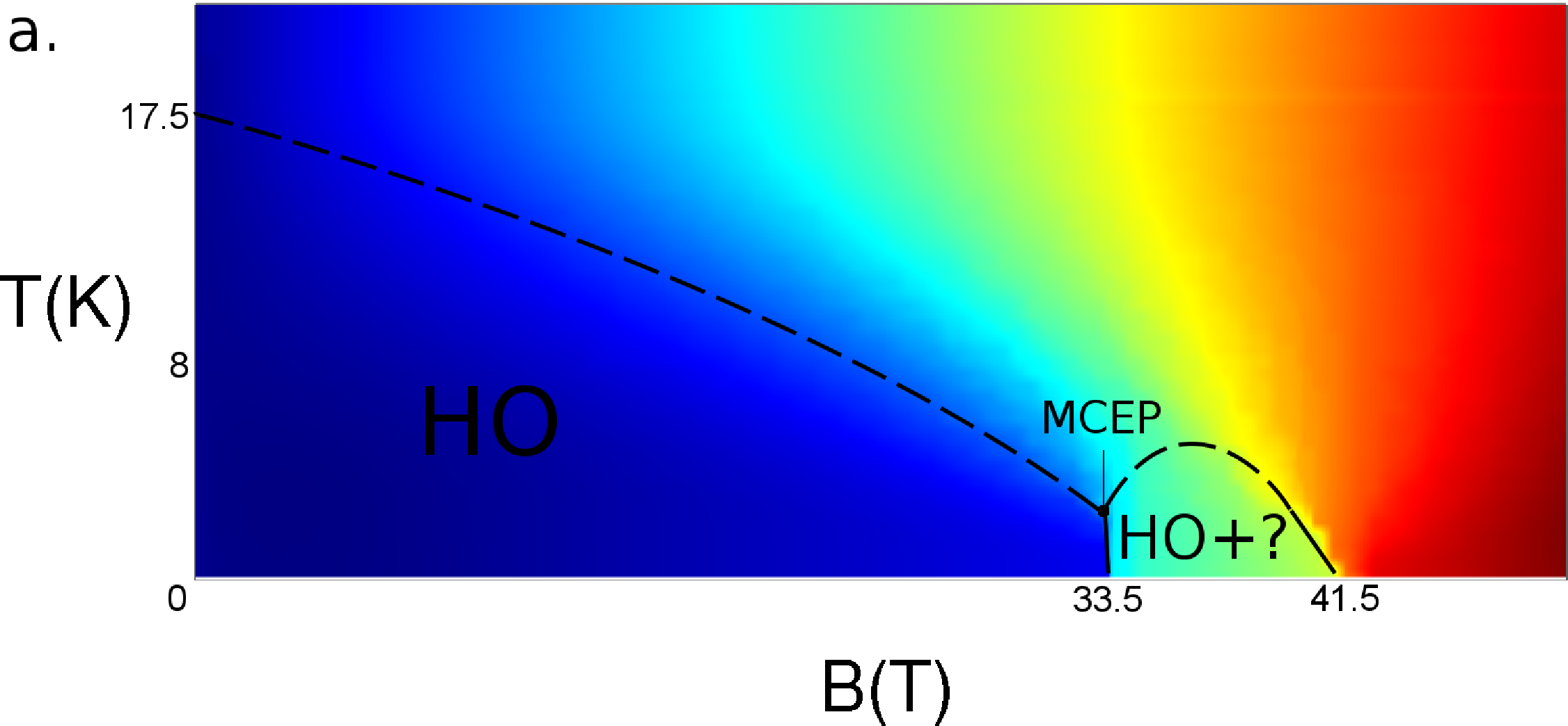}\centering
\hspace{0.05in}
\includegraphics[width=0.46\textwidth,height=0.34\textwidth]{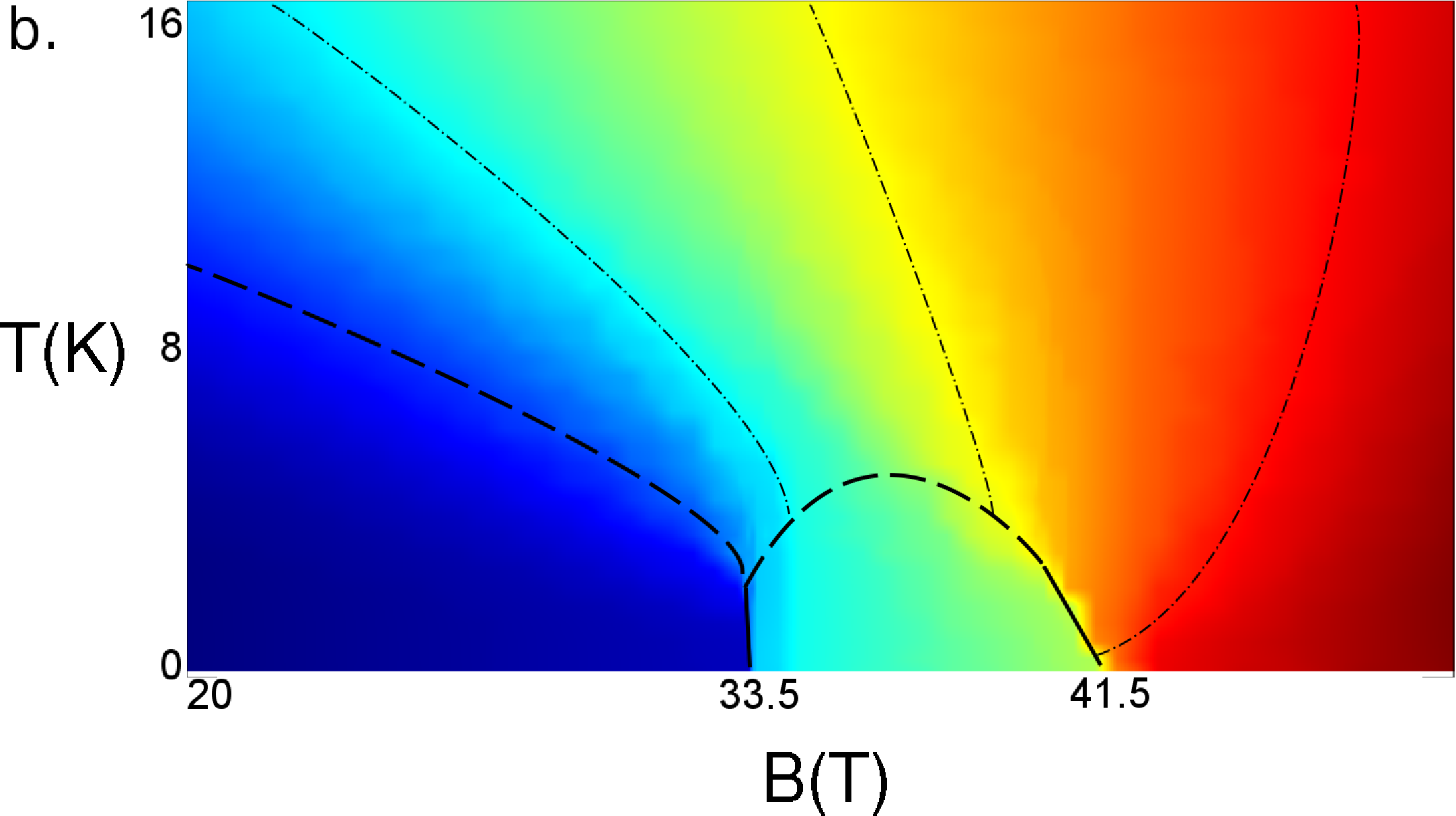}\centering
\vspace{0.07in}
\includegraphics[width=0.45\textwidth]{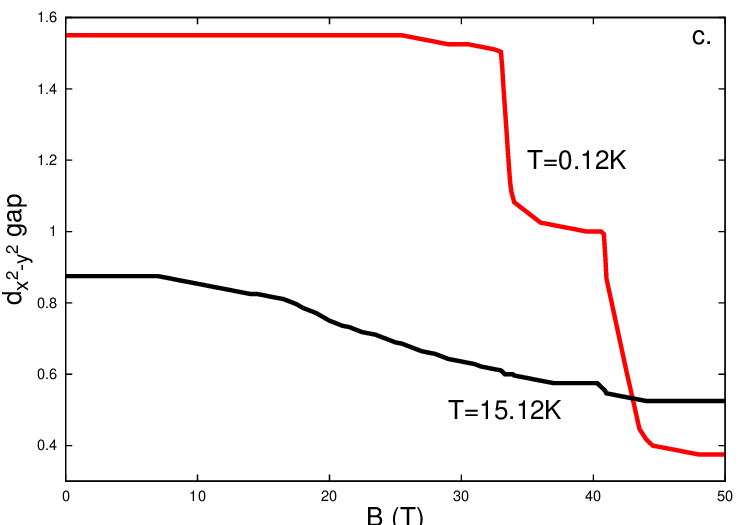}\centering
\hspace{0.05in}
\includegraphics[width=0.46\textwidth]{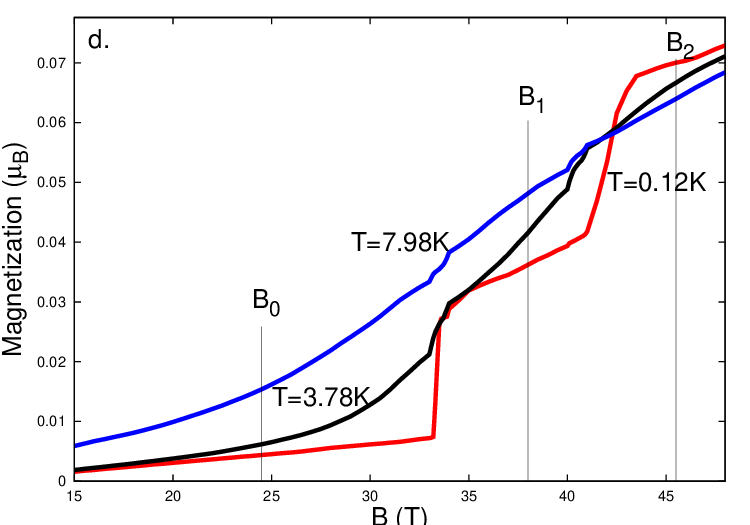}\centering
\vspace{0.07in}
\includegraphics[width=0.7\textwidth,height=0.35\textwidth]{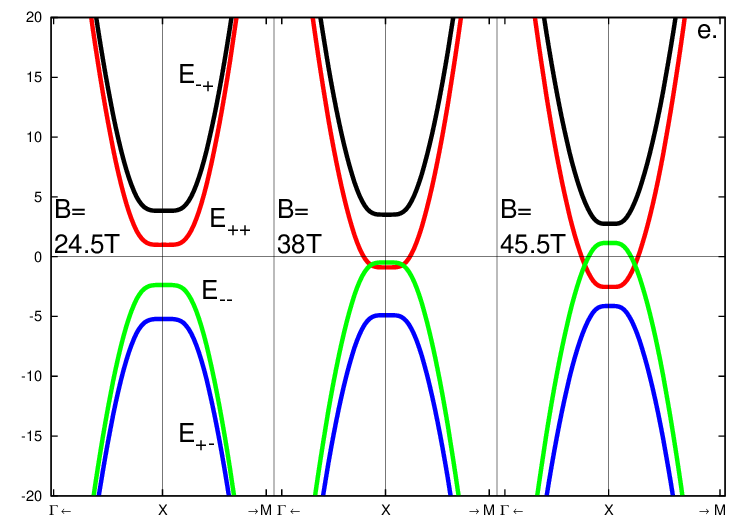}\centering
\end{minipage}
\caption{a. Self-consistently extracted ${\cal B}-T$ phase diagram of the chiral d-spin density wave. The chiral HO is present in the whole phase diagram. The reduced
magnitude of the order parameters in the high field and temperature regions, mask its presence providing a thermodynamic phase boundary (dashed lines). b. A Metamagnetic
Critical End Point (MCEP) is located at $T\simeq 3$K and ${\cal B}={\cal B}_{c1}=33.5$T. Two metamagnetic steps appear at ${\cal B}_{c1}=33.5$T and ${\cal B}_{c1}\simeq41$T
along with the formation of crossover areas (separated by dash-dotted lines), in agreement with the experimental observations \cite{reentrantHO,Kim}. c. For T$<$3K, the
$d_{x^2-y^2}$ order parameter is nearly constant until ${\cal B}={\cal B}_{c1}$, where it abruptly decreases. This is presumably the gap seen in neutron scattering experiments
\cite{Bourdarot}. Above the MCEP,  the gap evolves smoothly with the field, explaining the dependence arising from bulk measurements \cite{Mentink}. The saturation appearing
in high fields does not give rise to any thermodynamic anomaly, rendering the chiral HO experimentally undetectable. d. The metamagnetic steps are accompanied by an abrupt
increase of the magnetization. For higher tempe\-ra\-tu\-res the first order character becomes milder leading to a smooth evolution. e. The cascade of the metamagnetic
transitions, appears due to additional band crossings at each ${\cal B}_c$ \cite{AperisLT}. The reconstruction of the Fermi surface increases steeply the carriers of the
system and concomitantly the magnetization.}
\label{fig:1}
\end{figure}

The resulting ${\cal B}-T$ phase diagram is depicted in Fig.~\ref{fig:1}a,b. We observe that our phase diagram grasps to a large extent both qualitative and quantitative
features of the experimental observations \cite{reentrantHO,Kim}. At ${\cal B}=0$T, $\Delta_1=0$ and $\Delta_2=1.55{\rm meV}$. The latter order parameter corresponds to the
$id_{x^2-y^2}$ component which as we observe in Fig.~\ref{fig:1}c, slightly decreases upon raising the magnetic field for low temperatures. In contrast, the $d_{xy}$ order
pa\-ra\-me\-ter $\Delta_1$, rises monotonically. A MCEP \cite{Grigera} (MCEP) is located at ${\cal B}_{c1}=33.5$T and T$\simeq$3K, down to which the
driving HO gap, $\Delta_2$, evolves con\-ti\-nu\-ously with the magnetic field satisfying the quadratic relation $\Delta_2({\cal
B}_z)/\Delta_2(0)\simeq 1-({\cal B}_z\up/{\cal B}_{c1}\up)^2$. This behaviour has been verified in magnetoresistance measurements \cite{Mentink}. In stark contrast, $\Delta_2$
remains practically constant for temperatures below the MCEP which could explain the field independent energy scale witnessed in the inelastic magnetic response at $\bm{Q}_1$
\cite{Bourdarot}, in terms of the spin exciton scenario \cite{IncommensurateSpinResonance}. A continuous evolution with the field is not compatible with the first-order
character of the meta\-ma\-gnetic step. This property can settle the controversy concerning the observed field dependence of the HO.

Our self-consistent phase diagram exhibits double-step metamagnetism (Fig.~\ref{fig:1}d) at ${\cal B}_{c1}$=33.5T and ${\cal B}_{c2}\simeq$41T, below 3K in agreement with
experiments \cite{reentrantHO,Kim}. Note that the occurrence of metamagnetism in such a high field scale is also related to the appearance of the $d_{xy}$ component, which
leads to an enhancement of the critical temperature of the dominant $d_{x^2-y^2}$ order parameter (see Appendix). The gradual increment of the field shifts the four
quasiparticle bands through the combined orbital-Zeeman coupling (see Eq.\ref{cSDWpoles}). At each ${\cal B}_c$ an additional energy branch crosses the Fermi level 
(Fig.~\ref{fig:1}e) providing an abrupt increase to the density of carriers of the system and a concomitant steep enhancement of the magnetization \cite{AperisLT}. 
For higher temperatures the first order character of the metamagnetic steps becomes milder, leading to a smoother increase of the magnetization in this region of fields.

The pronounced similaritities presented in our phase diagram compared to Ref.~\cite{reentrantHO,Kim}, are accompanied by important ramifications arising from the finite
chirality of the order parameter. First and more importantly, the chiral HO is present in the whole diagram, due to the enhanced critical temperature, originating from the
orbital magnetic-field coupling. For magnetic fields ${\cal B}_{c1}<{\cal B}<{\cal B}_{c2}$, both order parameters of the chiral phase suffer a severe decrease in their
magnitude. Consequently, the chiral $d_{xy}+id_{x^2-y^2}$ phase is still present in this region but unavoidably masked due to the destruction of its topological rigidity. The
HO persists even for ${\cal B}\geq{\cal B}_{c2}$, behaving as a polarized paramagnetic metal.

Despite of the continuous presence of the chiral HO, the emergence of the MCEP may lead to the generation of novel phases at ${\cal B}_{c1}$, that could
coexist with the chiral HO, constituting phases ${\rm III,V}$ of Ref. \cite{Kim}. It has been shown that under quite generic conditions, the effect of a Zeeman field on a
charge density wave can lead to a field-induced SDW transition and vice versa \cite{AperisLT,AperisEPL}. In this sense, the magnetic field permits a number of
possible induced phases \cite{Patterns,GV} such as the chiral spin singlet d-density wave phase, studied in the context of high-Tc superconductors
\cite{PolarChiral,Meissner,ANEYakovenko,FICDDW,ChiralityNernst}, or even the spin singlet and triplet electron nematic phases
\cite{Varma,Yamase,Metzner,Fradkin,FradkinChiral}.

\section{Giant Nernst signal and anomalous thermoelectricity}

\begin{figure}[b]\centering
\begin{minipage}[b]{4.8in}
\includegraphics[width=0.85\textwidth]{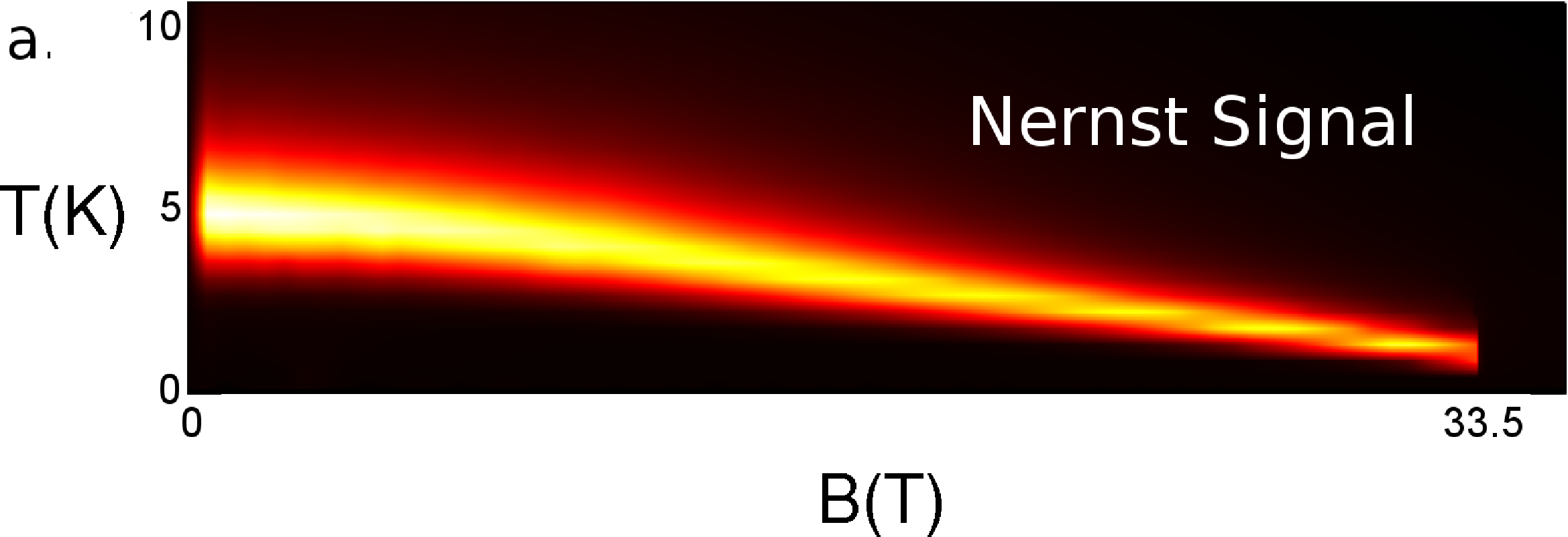}\\
\includegraphics[width=0.48\textwidth]{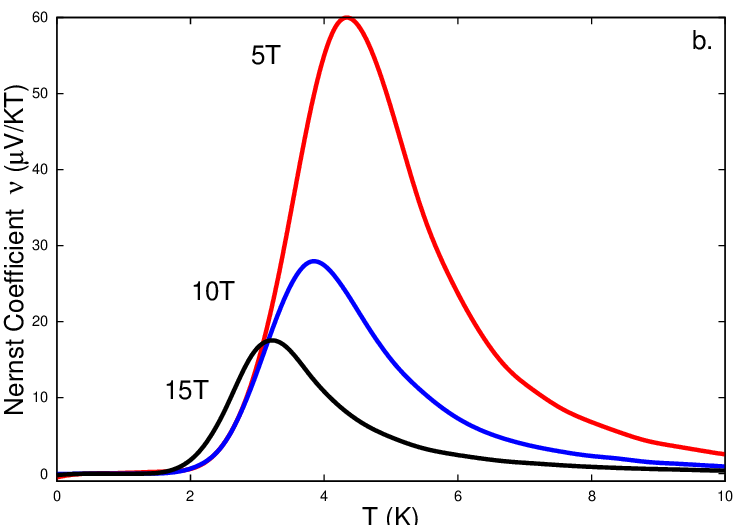}
\hspace{0.05in}
\includegraphics[width=0.48\textwidth]{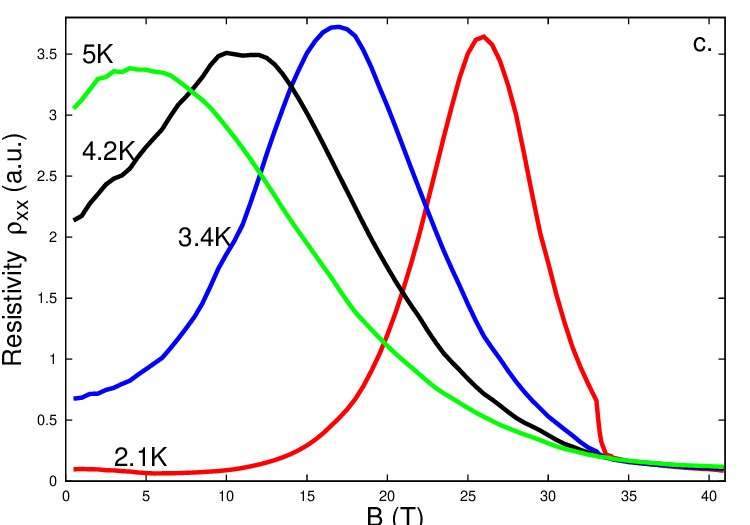}
\end{minipage}
\caption{ a. ${\cal B}-T$ dependence of the self-consistently calculated Nernst signal. The arising Nernst response presents a large positive signal, of about one order
larger than the expe\-ri\-mentally observed \cite{Bel}. The giant response originates from the topological robustness of the chiral phase \cite{ChiralityNernst}. The field
destruction of the Nernst signal at ${\cal B}_{c1}$ agrees with the experiment \cite{Behnia}, although here, it is accompanied by the dramatic reduction, rather than the
collapse of the $d_{xy},d_{x^2-y^2}$ order parameters. b. The Nernst coefficient $\nu=N/{\cal B}_z$ shows a `tilted-hill' profile, with a pronounced peak in the region 3-5K
following the experimental data \cite{Bel}. c. Field dependence of the resistivity for several temperatures. For $T<T_o$, the presence of a `kink', also found in experiments
\cite{Behnia}, may be naturally attributed to the field-enhancement of the $d_{xy}$ component.}
\label{fig:2}
\end{figure}

By taking into account both topological and quasiparticle contributions in computing the electric ($\sigma_{xx},\sigma_{xy}$) and thermoelectric conductivity
($\alpha_{xx},\alpha_{xy}$) tensor elements, we demonstrate that the Nernst response in the HO phase can be understood in terms of the chiral $d_{xy}+id_{x^2-y^2}$ HO
parameter. Using the values of the order parameters providing the phase diagram of Fig.~\ref{fig:1}, we obtain a giant Nernst signal, that has a magnitude of about one order
larger than the one observed in the experiments \cite{MatsudaLeiden}. The origin of the `tilted-hill' Nernst signal is directly related to the chirality and the strong Fermi
surface gapping. As presented in Ref.~\cite{ChiralityNernst}, the strong insulating character of the chiral HO phase forces the thermopower and the Nernst signal to become
equal at a thermoelectric crossing point. In the vi\-ci\-ni\-ty of this point, the usually large values of the thermopower also imply a Nernst signal of the same magnitude,
leading unavoidably to an enhancement of the latter. However, the thermoelectric rigitity and the tendency towards a giant Nernst signal are lost exactly at ${\cal B}_{c1}$.
The steep decrease of the $d_{x^2-y^2}$ gap across this first order transition and the concomitant Fermi surface reconstruction which partially destroys the insulating
properties of the phase, lead to the full supression of the response (Fig.~\ref{fig:2}a) in accordance with Ref.~\cite{Behnia}.

Furthermore, the Nernst coefficient exhibits a peaked structure with temperature and decreases upon raising the magnetic field already at 10K (Fig.~\ref{fig:2}b). Notice that
although this phase is also present even above 17.5K, it does not lead to a large Nernst signal, becoming in this manner experimentally elusive. We have also calculated the
resistivity for finite fields and temperatures. We observe in Fig.~\ref{fig:2}c that the field dependence of the resi\-sti\-vi\-ty exhibits a kink with the increase of the
magnetic field. This is a definite fingerprint of the field induced $d_{xy}$ component and coincides with the experimental findings \cite{Behnia}. This kink shifts to lower
magnetic fields upon raising temperature, since the $d_{xy}$ gap gradually weakens.

\section{Prediction of a non-linear ${\cal B}$-dependence for the Kerr angle due to the induced chirality}

\begin{figure}[h]\centering
\includegraphics[width=0.56\textwidth]{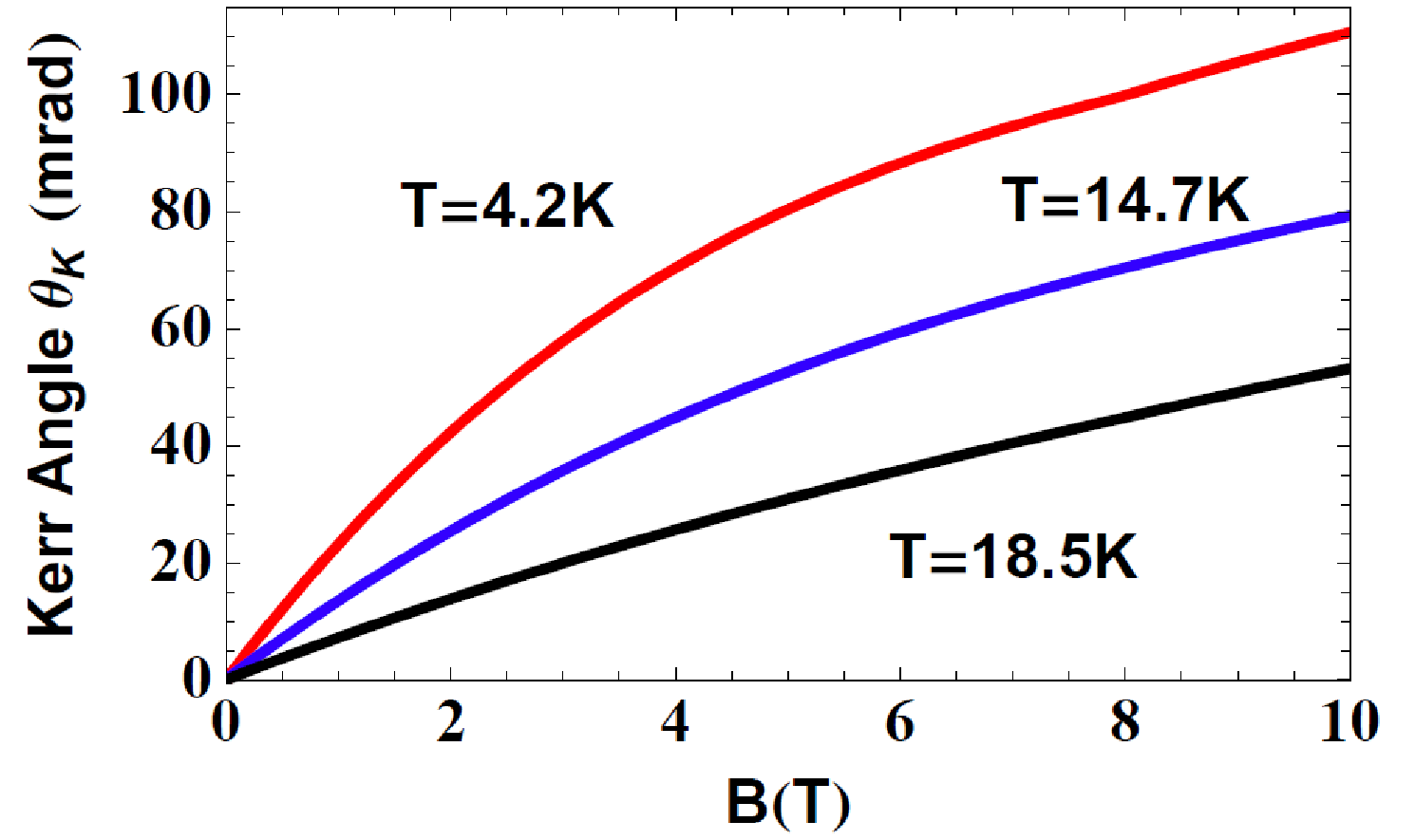}
\caption{Predicted magnetic field dependence of the Kerr angle due to the magnetic induction of the chiral HO. The emergence of the chiral d-SDW yields a large field-induced
Kerr angle. The latter can be observed only for finite magnetic fields, and even above $T_o$, due to the increase of the HO critical temperature arising from the magnetic
induction of the $d_{xy}$ order parameter (see Appendix). Starkingly, for low temperatures we obtain a characteristic \textit{non-linear} ${\cal B}$-dependence which can
identify the emergence of the field-induced chiral HO proposed here.}
\label{fig:3}
\end{figure}

To verify our scenario we propose the measurement of the Kerr angle \cite{PolarChiral} in the presence of a static magnetic field. This type of measurement is an ideal
probe of the possible emergence of chirality in the HO phase \cite{Kapitulnik}. The field-induced chiral HO will provide the dominant contribution to the
Kerr angle. More importantly, we expect that the a\-ri\-sing Kerr angle $\vartheta_K\up$ will exhibit a \textit{distinctive} magnetic field dependence when the sample is
cooled down and heated up again in the presence of a static external magnetic field (Fig.~\ref{fig:3}). The Kerr angle \cite{Kapitulnik,PolarChiral,MineevKerr} is defined as
\bea\vartheta_K\up=\frac{4\pi}{n(n^2-1)da^2}\frac{\sigma_{xy}^{\Im}(\omega)}{\omega}\,,\eea

\noi where $n$ is the refractive index, $d$ is the c-axis thickness of the sample, $a=5{\AA}$ the in plane lattice constant and $\sigma_{xy}^{\Im}$ the imaginary part of the
dynamical Hall conductivity. In the low field regime (${\cal B}\leq 20T$), the dynamical Hall conductivity is dominated by the chirality contribution. For a Sagnac
inteferometer \cite{Kapitulnik} $\hbar\omega=0.8eV=6420{\rm cm^{-1}}$ and $\lambda=1550{\rm nm}$, while the plasma frequency of ${\rm URu_2Si_2}$ is $\hbar\omega_p=2131{\rm
cm^{-1}}\simeq 266 {\rm meV}$ \cite{Infrared}. The refractive index in this case is defined as $n=\sqrt{1-\left(\omega_p/\omega\right)^2}$. Since $\omega>>\omega_p$ we may use
the approximation $n\simeq 1$ and $n^2-1=-\left(\omega_p/\omega\right)^2$. We find that the Kerr angle is equal to
\bea
\vartheta_K\up({\cal B},T)
&=&\frac{4\pi}{n(n^2-1)da^2}\frac{\sigma_{xy}^{\Im}(\omega)}{\omega}\simeq-
\frac{2}{n(n^2-1)}\frac{e^2}{\hbar c}\frac{\lambda}{d}\frac{\Delta_1({\cal B},T)}{\hbar\omega}
\no\\
&=&-2\frac{e^2}{\hbar c}\frac{\lambda}{d}\frac{\Delta_1({\cal B},T)}{\hbar\omega_p}\left(\frac{\omega}{\omega_p}\right)\,.
\eea

\noi By taking into account the value of the fine structure constant $e^2/\hbar c=1/137$ and by setting $d\simeq 2{\rm nm}$, we obtain a rough estimation of the Kerr angle as
$\vartheta_K\up\simeq 128 \Delta_1 {\rm mrad}$ with $\Delta_1$ in ${\rm meV}$. The maximum value of $\Delta_1$ is equal to $1.5{\rm meV}$, which defines an upper bound
$\vartheta_K\up\leq190{\rm mrad}$. The magnitude of the Kerr angle demonstrates that the expected response is very large and easily detectable. The observation of the
particular \textit{non-linear} ${\cal B}$-dependent chirality contribution to the Kerr angle, as opposed to the linearly dependent quasiparticle contribution, constitutes a
sharp signature of our magnetic-field-induced chiral HO scenario, unveiling its possible connection to the intricate ${\rm URu_2Si_2}$ phenomenology.

\section{Summary and conclusions}

In conclusion, assuming that the zero-field HO in ${\rm URu_2Si_2}$ is an unconventional $id_{x^2-y^2}$ spin density wave phase that preserves TRS, we have demonstrated
that both the giant anomalous Nernst signal and the double-step metamagnetism observed in the HO of this material can be understood in a unified way, in terms of a magnetic
field-induced $d_{xy}$ component which renders the HO \textit{chiral} in finite magnetic fields. This chiral HO is characterized by a finite orbital ferromagnetic moment
(arising from a finite Berry curvature) which ensures a robust giant anomalous Nernst signal. At the same time, the interplay of orbital and Zeeman moments allow for a
sequential softening of the emergent four quasiparticle branches by the magnetic field, leading to double-step metamagnetism. 

Within a mean-field theory for the magnetic field-induced chiral HO, while taking fully into account effects of induced orbital moment in the chiral phase, we have numerically
obtained the ${\cal B}-T$ phase diagram of this material. For the same set of numerical self-consistent data, we have shown  that also the temperature profile of the obtained
Nernst signal matches that of the relevant experiments. In order to unveil the chiral nature of the HO, we propose a Kerr effect experiment in the presence of an external
magnetic field. We predict a characteristic temperature and magnetic field profile of the Kerr angle. If the latter is indeed observed, this would open a new perspective to
our understanding of the HO in ${\rm URu_2Si_2}$.

\section*{Acknowledgements}
We are indebted to J. Mydosh, P. M. Oppeneer, Y. Matsuda, K. Behnia and A. Kapitulnik for illuminating discussions. This work was partially funded by ${\rm \Pi EVE}$ by NTU
Athens. A. A. wishes to thank the organizers of the Workshop on Hidden Order, Superconductivity and Magnetism in ${\rm URu_2Si_2}$, Lorentz Center, Leiden for their
hospitality.

\newpage

\appendices

\section{Mean-field decoupling and order parameter symmetries}

Based on the phenomenology of ${\rm URu_2Si_2}$, we assume that in the real material, the chiral d-SDW orders at the incommensurate wave vector $\bm{Q}_1=(1\pm 0.4,0,0)$.
However, describing microscopically a chiral order parameter with this ordering wave-vector, demands taking into account the main four Fermi lines that give rise to this
nesting symmetry \cite{Elgazzar,OppeneerEtal}. This is a very complex and unnecessary task since our results do not depend on the microscopic details of the model. Our
results concerning the anomalous Hall thermoelectric transport are based solely on the topological content and the symmetry properties of this phase. Moreover, the phase
diagram and the metamagnetic transitions are also based on the general principles of itinerant metamagnetism, according to which this phenomenon ori\-gi\-na\-tes from band
crossing \cite{BandCrossing}. Under these conditions it is permissible to consider a simplified single-band tight-binding model characterized by the nesting condition
$\varepsilon(\bm{k}+\bm{Q})=-\varepsilon(\bm{k})=-[-2t(\cos k_x+\cos k_y)]$ with the commensurate nesting wave-vector $\bm{Q}=(\pi,\pi)$. This condition, should stand in a
similar way for the real bands of URu$_2$Si$_2$, and within our approximate simple model we simulate this situation. Finally, within this framework we consider
the $d$-wave harmonics $\Delta_1\up(\bm{k})=\Delta_1\up\sin k_x\sin k_y$, $\Delta_2\up(\bm{k})=\Delta_2\up(\cos k_x-\cos k_y)$ with $\Delta_{1,2}\up=\Delta^*_{1,2}$.

To study the phase diagram of the chiral d-SDW phase we consider that it is driven by an a inter-site Coulomb repulsion including nearest and next-nearest neighbours
\bea{\cal V}_{int}\up&=&-\frac{1}{v}\sum_{\bm{k},\bm{k}'}\sum_{s,s'}V(\bm{k}-\bm{k}')
c_{\bm{k},s}^{\dag}c_{\bm{k}+\bm{Q},s'}\up c_{\bm{k}'+\bm{Q},s'}^{\dag}c_{\bm{k}',s}\up\ph\eea

\noi with $s,s'=\uparrow,\downarrow$ and $V(\bm{q})=2V'(\cos q_x+\cos q_y)+4V''\cos q_x\cos q_y$. The momenta $\bm{k},\bm{k}'$ belong to the whole Brillouin zone (${\cal
B.Z.}$) if $\bm{Q}$ is incommensurate and in the reduced Brillouin zone if $\bm{Q}$ is commensurate satisfying $\bm{k}+2\bm{Q}=\bm{k}$. The potential is separable providing
$V(\bm{k}-\bm{k}')=\sum_{n}V_n\up f_n\up({\bm{k}})f_n\up({\bm{k}'})$ with $n=1,2,..,8$ corresponding to the form factors $f_n(\bm{k}):\cos k_x\pm\cos k_y$, $\sin k_x\pm\sin
k_y$, $\cos k_x\cos k_y$, $\sin k_x\sin k_y$, $\cos k_x\sin k_y$, $\sin k_x\cos k_y$ with the driving potentials $V_n\up=V'$ for $n=1,2,3,4$ and $V_n\up=4V''$ for the rest.
Then within a mean-field treatment we introduce the staggered order parameters
\bea \Delta_{\bm{Q}}^{ss}(\bm{k})&=&-\frac{1}{v}\sum_{\bm{k}'}V(\bm{k}-\bm{k}')\left<c_{\bm{k}'+\bm{Q},s}^{\dag}c_{\bm{k}',s}\up\right>
=\sum_n\Delta_{\bm{Q},n}^{ss}(\bm{k})\quad s=\uparrow,\downarrow
\eea

\noi and the corresponding irreducible $\bm{k}$-independent staggered order parameters

\bea \Delta_{\bm{Q},n}^{ss}&=&-\frac{1}{v}\sum_{\bm{k}}V_n\up f_n\up(\bm{k})\left<c_{\bm{k}+\bm{Q},s}^{\dag}c_{\bm{k},s}\up\right>\quad s=\uparrow,\downarrow\,.\eea

\noi The interaction decouples in the following way
\bea{\cal V}_{int}\up&=&\sum_{\bm{k}}\left\{\Delta_{\bm{Q}}^{\uparrow\uparrow}(\bm{k})c_{\bm{k},\uparrow}^{\dag}c_{\bm{k}+\bm{Q},\uparrow}\up+\Delta_{\bm{Q}}^{
\downarrow\downarrow}(\bm{ k})c_{\bm{k },\downarrow}^{\dag}c_{\bm{k}+\bm{Q},\downarrow}\up+h.c.\right\}+
v\sum_{n,s}\frac{|\Delta_{\bm{Q},n}^{ss}|^2}{V_n\up}\,,\qquad\eea

\noi with $v$ the volume of the corresponding ${\cal B.Z.}$. By introducing the spinor $\Psi_{\bm{k}}^{\dag}=(c_{\bm{k},\uparrow}^{\dag}\phd c_{\bm{k},\downarrow}^{\dag}\phd
c_{\bm{k}+\bm{Q},\uparrow}^{\dag}\phd c_{\bm{k}+\bm{Q},\downarrow}^{\dag})$ and by using the isospin and spin Pauli matrices $\tau_i, s_i$ with $i=1,2,3$  complemented by
the related unit matrices $\tau_0\up, s_0\up$, we furnish a representation of the $4\times 4$ mean-field hamiltonian as Kronecker pro\-ducts of the form
$\tau_{\mu}\otimes s_{\nu}\up$ where $\mu,\nu=0,1,2,3$ (for simplicity we omit $\otimes$). The most general chiral d-SDW state is defined as
$\Delta_{\bm{Q}}^z(\bm{k})=\Delta_1\up(\bm{k})-i\Delta_2\up(\bm{k})$. Moreover we introduce the kinetic term of the paramagnetic state ${\cal
H}_0\up=\sum_{\bm{k},s}\left\{(\varepsilon(\bm{k})-\mu) c_{\bm{k},s}^{\dag}c_{\bm{k},s}\up+(\varepsilon(\bm{k}+\bm{Q})-\mu)c_{\bm{k}+\bm{Q},s}^{\dag}c_{\bm{k}+\bm{Q},s}
\up\right\}$ along with a chemical potential $\mu$. Putting together all these terms and using the fact that in the spin-triplet state
$\Delta_{\bm{Q}}^z=\Delta_{\bm{Q}}^{\uparrow\uparrow}=-\Delta_{\bm{Q}}^{\downarrow\downarrow}$, we obtain the quasiparticle Hamiltonian
\bea\no{\cal H}_{q-p}\up
&=&\sum_{\bm{k}}\Psi_{\bm{k}}^{\dag}\Bigl\{\frac{\Delta_1\up(\bm{k})+\Delta_1^*(\bm{k})}{2}\tau_1\up s_3\up+\frac{\Delta_1\up(\bm{k})-\Delta_1^*(\bm{k})}{2}
i\tau_2\up s_3\up\\\no
&-&i\frac{\Delta_2(\bm{k})-\Delta_2^*(\bm{k})}{2}\tau_1\up s_3\up+\frac{\Delta_2(\bm{k})+\Delta_2^*(\bm{k})}{2}\tau_2\up s_3\up
\\&+&\frac{\varepsilon(\bm{k})+\varepsilon(\bm{k}+\bm{Q})}{2}\tau_0\up s_0\up+\frac{\varepsilon(\bm{k})-\varepsilon(\bm{k}+\bm{Q})}{2}
\tau_3\up s_0\up-\mu\tau_0\up s_0\up\Bigl\}\Psi_{\bm{k}}\up\,,\eea

\noi and the elastic energy needed to build up the two density wave gaps
\bea {\cal H}_{\bm{\Delta}^2}\up=2v\left(\frac{|\Delta_1|^2}{4V''}+\frac{|\Delta_2|^2}{V'}\right)\,.\eea

\section{Intrinsic orbital moment and magnetic field coupling}

With these definitions we may write the quasiparticle hamiltonian in the condensed form
${\cal H}_{q-p}\up=\sum_{\bm{k}}\Psi_{\bm{k}}^{\dag}\left\{\widehat{\bm{g}}(\bm{k})\cdot\bm{\tau}-\mu\right\}\Psi_{\bm{k}}\up$ where we have introduced  the vector
$\widehat{\bm{g}}(\bm{k})=\left(\Delta_1(\bm{k}) s_3\up,\Delta_2(\bm{k}) s_3\up,\varepsilon(\bm{k}) s_0\up\right)$. Diagonalizing the above hamiltonian yields the
eigenstates $|\Phi_{s,\nu}(\bm{k})\rangle$ with energy dispersions $E_{s,\nu}(\bm{k})=-\mu+\nu E(\bm{k})$ and $\nu=\pm,s=\uparrow,\downarrow$,
$E(\bm{k})=|\widehat{\bm{g}}(\bm{k})|=\sqrt{\varepsilon^2(\bm{k})+|\Delta_{\bm{Q}}^z(\bm{k})|^2}$. The non trivial topological content of this Hamiltonian generates a non zero
$U(1)$ Berry connection \cite{NiuReview} $\bm{{\cal A}}_{s,\nu}(\bm{k})
=\langle \Phi_{s,\nu}(\bm{k})\mid i\bm{\nabla}_{\bm{k}}\up\mid \Phi_{s,\nu}(\bm{k})\rangle$ and a corresponding Berry curvature defined as
$\bm{\Omega}_{s,\nu}(\bm{k})=\bm{\nabla}_{\bm{k}}\times\bm{{\cal A}}_{s,\nu}(\bm{k})=\Omega^z_{s,\nu}(\bm{k})\bm{\hat{z}}$, i.e. oriented along the $z$-axis. The detailed
expression for the Berry curvature reads \cite{FieldInduced,Chirality,ANE,Goswami}
\bea\Omega_{s,\nu}^z(\bm{k})=\frac{1}{2}Tr_{ s}\up\left\{\frac{-\nu a^2}{2E^3(\bm{k})}\widehat{\bm{g}}(\bm{k})\cdot\left(\frac{\partial\widehat{\bm{g}}(\bm{k})}{\partial
k_x\up}\times\frac{\partial\widehat{\bm{g}}(\bm{k})}{\partial k_y\up}\right)\right\}\,.\eea

\noi We observe that the Berry curvature is not depending on spin while it just changes sign when $\nu=\pm$. The intrinsic orbital moment of this state is
straighforward calculated using the definition \cite{NiuReview}
\bea\no \bm{m}_{s,\nu}\up(\bm{k})&=&\frac{e}{2\hbar
i}\langle\bm{\nabla}_{\bm{k}}\Phi_{s,\nu}\up(\bm{k})\mid\times\left[{\cal
H}(\bm{k})-E_{s,\nu}\up(\bm{k})\right]\mid\bm{\nabla}_{\bm{k}}\Phi_{s,\nu}\up(\bm{k})\rangle\\
&=&\frac{e\nu}{\hbar}
E(\bm{k})\bm{\Omega}_{s,\nu}\up(\bm{k})\,.\eea

\noi One may observe that the orbital moment is band independent.  When we apply a magnetic field, the system interacts with both the Zeeman and the orbital moment
$m_z(\bm{k})=eE(\bm{k})\Omega_{s,+}^z(\bm{k})/\hbar$ yielding the following interacting Hamiltonian ${\cal H_{{\cal
B}}}(\bm{k})=-(\mu_B\tau_0\up s_3\up-m_z(\bm{k})\tau_0\up s_0\up){\cal B}$. The inclusion of the magnetic coupling leads to the four field dependent eigenenergies:
\begin{eqnarray}\label{cSDWpoles}
E_{s,\nu}^{{\cal B}}(\bm{k})=-[s\mu_B-m_z(\bm{k})]{\cal B} +\nu \sqrt{\varepsilon^2(\bm{k})+|\Delta_{\bm{Q}}^z(\bm{k})|^2}\ \ ,\ \ \nu,s=\pm
\end{eqnarray}

\noi corresponding to the conduction and valence bands of the upper and lower spin sectors respectively.

\section{Free energy and self-consistency equations}

Having set up the microscopic description of the chiral d-SDW condensate, inte\-racting with both Zeeman and the intrisic orbital moments with a magnetic field, we may 
solve the self-consistency equations that derive from the minimization of the free-energy functional
\bea{\cal F}=2v\left(\frac{\Delta_1^2}{4V''}+\frac{\Delta_2^2}{V'}\right)-\frac{1}{\beta}\sum_{\bm{k},s,\nu}
\ln\left(1+e^{-\beta E_{s,\nu}^{{\cal B}}(\bm{k})}\right)\,,\eea

\noi with $t=50{\rm meV}$, $\mu=0.69{\rm meV}$, $\mu_B\up=0.058{\rm meV/T}$, $a=5{\AA}$, $V'=23.5 {\rm meV}$ and $V''=35.25 {\rm meV}$. Minimization of the above yields
the following two self-consistency equations, that determine the order parameters of the chiral d-SDW \cite{FieldInduced}
\bea\no
\Delta_i\up=\frac{-V_{i}\up}{2v}\sum_{\bm{k},s,\nu}\left\{\nu
\Delta_i\up\frac{f_i^2(\bm{k})}{2E(\bm{k})}+\frac{ea^2t}{\hbar}{\cal
B}_z\up\widetilde{\Delta}_{i}\up
\frac{S(\bm{k})}{E^2(\bm{k})}\left[1-2\left(\frac{\Delta_i\up(\bm{k})}{E(\bm{k})}\right)^2\right]\right\}
\\\times n_F\up\left[E_{s,\nu}^{\bm{{\cal
B}}}(\bm{k})\right]\,,\eea

\noi with $V_1\up=4V''$, $V_2\up=V'$, $\widetilde{\Delta}_{1,2}\up=\Delta_{2,1}\up$, $S(\bm{k})=\sin^2 k_x\cos^2 k_y+\sin^2 k_y$ and $n_F\up$ the Fermi-Dirac distribution.

\section{Landau theory of the magnetic-field-induced chiral d-spin density wave HO and T$_c$ enhancement}

In order to understand phenomenologically the mechanism leading to the field-induced chiral HO as this arises microscopically from the previous appendices, we can construct a
employ a Landau description. The numerical solution of the equations, points to the following Landau picture for the chiral d-spin density wave order

\bea{\cal F}=\alpha_1\up\frac{\Delta_1^2}{2}+\alpha_2(T-T_o\up)\frac{\Delta_2^2}{2}+\beta\frac{\Delta_2^4}{4}-g\Delta_1\up\Delta_2\up{\cal B}_z\up\eea

\noi The phenomenological constants $\alpha_{1,2}\up,\beta$ and the field ${\cal B}_z\up$, are positive. The coefficient $\alpha_1\up$ ensures that the $d_{xy}\up$ component
is magnetic-field induced. From the numerical analysis we have discerned that the sign of the orbital coupling $g$, is also positive. With this convention $\Delta_{1,2}\up$
have the same sign. The mean-field solution for the $\Delta_1$ component yields

\bea 
\frac{\partial{\cal F}}{\partial\Delta_1}=0\Rightarrow\alpha_1\Delta_2\up&=&g\Delta_2\up{\cal B}_z\,.\eea

\noi One may `integrate out' the induced $\Delta_1$ component and obtain an effective theory only in terms of the driving $d_{x^2-y^2}$ order parameter $\Delta_2$. This is
straightforward, yielding

\bea{\cal F}_{eff}=\alpha(T-T_o\up)\frac{\Delta_2^2}{2}+\beta\frac{\Delta_2^4}{4}-\frac{g^2}{\alpha_1}\Delta_2^2{\cal B}_z^2=\alpha\left[T-T_o\up({\cal
B}_z\up)\right]\frac{\Delta_2^2}{2}+\beta\frac{\Delta_2^4}{4}\,.\eea

\noi Note, that since $\alpha_1>0$, the quadratic magnetic-field-coupling leads to an enhanced critical temperature $T_o\up({\cal B}_z\up)=T_o\up+\frac{g^2}{\alpha_1}{\cal
B}_z^2$ for the $d_{x^2-y^2}$ component and consequently for the chiral order. In the magnetic-field-induced chiral d-spin density wave, the $\Delta_2$ order parameter is
driving the transition.

\section{Magnetization and thermoelectric response}

As far as the magnetic and transport properties are concerned, we have used the expressions found in the theory of orbital magnetization \cite{NiuReview}. The
magnetization is the summation of both spin and orbital contributions in the following way ${\cal M}_z^{\cal B}={\cal M}_{z,spin}^{\cal B}+{\cal M}_{z,orb}^{\cal B}$ where:
\bea{\cal M}_{z,spin}^{\cal B}&=&\frac{1}{v}\sum_{\bm{k},s,\nu}\left\{\left(1+\frac{e{\cal B}a^2}{\hbar}\Omega_{s,\nu}^z(\bm{k})\right)\mu_B\up s\ph n_F\up[E_{s,\nu}^{\cal
B}(\bm{k})]\right\}\\\no
{\cal M}_{z,orb}^{\cal B}&=&\frac{1}{v}\sum_{\bm{k},s,\nu}\Biggl\{\left(1+\frac{e{\cal B}a^2}{\hbar}\Omega_{s,\nu}^z(\bm{k})\right)m_z(\bm{k})n_F\up[E_{s,\nu}^{\cal
B}(\bm{k})]\\
&+&\frac{ea^2}{\hbar}\Omega_{s,\nu}^z(\bm{k})k_BT\ln\left(1+e^{-E_{s,\nu}^{\cal B}(\bm{k})/k_BT}\right)\Biggl\}\,.\eea

\noi The above expressions include the Berry phase correction to the electronic density of states, as this is apparent from the factor $\left(1+\frac{e{\cal
B}a^2}{\hbar}\Omega_{s,\nu}^z(\bm{k})\right)$ in the first equation.\\

For transport properties, we have taken into account both topological $(\sigma_{xy,top}^{\cal B}$, $\alpha_{xy,top}^{\cal B})$ and quasiparticle
contributions ($\sigma_{xy,q-p}^{\cal B}$, $\alpha_{xy,q-p}^{\cal B})$ in calculating the Hall transport, while we have incorporated the Berry phase corrections in the
scattering time and density states of the system. We have used $\tau_{s,\nu}(\bm{k})\equiv\tau=3.3\cdot 10^{-13}s$. Based on experimental data \cite{Measson} $m^*=25m_e$, and
$\omega_c\tau({\cal B}=1T)=0.08$, the estimated scattering time is of the order of $10^{-12}s$. In order to illustrate the chirality induced giant Nernst signal, we have on
purpose selected a 3 times smaller re\-la\-xa\-tion time so to demonstrate that the possible enhan\-cement of the quasi\-particle contri\-bution due to a large $\tau$, is not
crucial for the anomalous Hall thermo\-elec\-tricity. In finite magnetic fields, the initially constant relaxation time, becomes field and momentum dependent in the following
manner: $\tau_{s,\nu}^{\cal B}(\bm{k})=\frac{\tau}{1+\frac{e}{\hbar}{\cal B}\Omega_{s,\nu}^z(\bm{k})}$. The electric and thermoelectric coefficients read:
\bea\no
\sigma_{xx}^{\cal B}
&=&-\frac{1}{v}\sum_{\bm{k},s,\nu}\frac{n_F'\left[E_{s,\nu}^{\cal B}({\bm{k}})\right]}{1+\frac{e}{\hbar}{\cal B}\Omega_{s,\nu}^z(\bm{k})}\left(\frac{\tau_{s,\nu}^{\cal
B}(\bm{k})}{\hbar}\right)\left[v_{s,\nu}^{x,\cal B}(\bm{k})\right]^2\,,\\\no
\alpha_{xx}^{\cal B}
&=&+\frac{1}{v}\sum_{\bm{k},s,\nu}\frac{n_F'\left[E_{s,\nu}^{\cal B}({\bm{k}})\right]}{1+\frac{e}{\hbar}{\cal B}\Omega_{s,\nu}^z(\bm{k})}\left(\frac{\tau_{s,\nu}^{\cal
B}(\bm{k})}{\hbar}\right)\left[v_{s,\nu}^{x,\cal B}(\bm{k})\right]^2\left(\frac{E_{s,\nu}^{\cal B}(\bm{k})}{k_B\up T}\right)\,,\\\no
\sigma_{xy,top}^{\cal B}&=&-\frac{1}{v}\sum_{\bm{k},s,\nu}
n_F\up[E_{s,\nu}^{\cal B}({\bm{k}})]\Omega_{s,\nu}^z(\bm{k})\,,\\\no
\alpha_{xy,top}^{\cal B}&=&+\frac{1}{v}\sum_{\bm{k},s,\nu}
\left\{\left(\frac{E_{s,\nu}^{\cal B}(\bm{k})}{k_B\up T}\right)n_F\up[E_{s,\nu}^{\cal B}(\bm{k})]+\ln\left(1+e^{-E_{s,\nu}^{\cal B}(\bm{k})/k_B\up
T}\right)\right\}\Omega_{s,\nu}^z(\bm{k})\,,\\\no
\sigma_{xy,q-p}^{\cal B}&=&-\frac{1}{v}\sum_{\bm{k},s,\nu}
\frac{n_F'\left[E_{s,\nu}^{\cal B}({\bm{k}})\right]}{1+\frac{e}{\hbar}{\cal B}\Omega_{s,\nu}^z(\bm{k})}\left(\frac{\tau_{s,\nu}^{\cal
B}(\bm{k})}{\hbar}\right)^2\frac{ea^2{\cal B}}{\hbar}v_{s,\nu}^{x,{\cal B}}(\bm{k})\left(\varepsilon_{ijz}v_{s,\nu}^{j,\cal B}(\bm{k})\frac{\partial}{\partial
k_i}\right)\\\no
&\times&v_{s,\nu}^{y,\cal B}(\bm{k}),i,j=x,y\,,\\\no
\alpha_{xy,q-p}^{\cal B}&=&-\frac{1}{v}\sum_{\bm{k},s,\nu}
\frac{n_F'\left[E_{s,\nu}^{\cal B}({\bm{k}})\right]}{1+\frac{e}{\hbar}{\cal B}\Omega_{s,\nu}^z(\bm{k})}\left(\frac{\tau_{s,\nu}^{\cal
B}(\bm{k})}{\hbar}\right)^2\frac{ea^2{\cal B}}{\hbar}v_{s,\nu}^{x,\cal B}(\bm{k})\left(\varepsilon_{ijz}v_{s,\nu}^{j,\cal B
}(\bm{k})\frac{\partial}{\partial k_i}\right)\\\no
&\times& v_{s,\nu}^{y,\cal B}(\bm{k})
\left(\frac{E_{s,\nu}^{\cal B}(\bm{k})}{k_B\up T}\right),i,j=x,y\,.\quad
\eea

\noi with units $[\sigma]=ea^2/\hbar$ and $[\alpha]=(k_B\up/e)[\sigma]$. We have also introduced the field dependent velocities $\bm{v}_{s,\nu}^{\cal
B}(\bm{k})=\bm{\nabla_k}\up E_{s,\nu}^{\cal B}(\bm{k})$. The Nernst signal and the resistivity are calculated from the following expressions
\bea{\cal N}=\frac{E_y\up}{-\partial_x\up T}=\frac{\sigma_{xx}^{\cal B}\alpha_{xy}^{\cal B}-\alpha_{xx}^{\cal B}\sigma_{xy}^{\cal B}}{\sigma_{xx}^{\cal B}\sigma_{yy}^{\cal
B}+\sigma_{xy}^{\cal B}\sigma_{xy}^{\cal B}}\,\,,\qquad\qquad
\rho_{xx}^{\cal B}=\frac{\sigma_{xx}^{\cal B}}{\sigma_{xx}^{\cal B}\sigma_{yy}^{\cal B}+\sigma_{xy}^{\cal B}\sigma_{xy}^{\cal B}}\,.\eea

\section{Kerr effect in the presence of a static magnetic field}

\noi For the calculations of the Kerr angle in the presence of small external static magnetic fields, only the topological dynamical Hall conductivity is important in our
case, defined as $ \sigma_{xy}\up(\omega)=\frac{\Pi^{xy}_{cc}(\bm{q}\rightarrow
0,i\omega_s\up\rightarrow\omega+i0^+)}{-i\omega}$ with $\Pi_{xy}^{cc}(\bm{q},i\omega_s\up)$ the current-current polarization function originating from the bubble diagram:
\bea\no
\Pi_{ij}^{cc}(i\omega_s\up,\bm{q})=\frac{a^2}{\beta}\sum_{ik_n\up,\bm{k}}Tr_{\tau, s}\up\left\{\widehat{{\cal
G}}(ik_n\up,\bm{k})\ph\widehat{\Gamma}_i^{\mu}(\bm{k},\bm{k}+\bm{q})\tau_{\mu}\ph
\widehat{{\cal
G}}(ik_n\up+i\omega_s\up,\bm{k}+\bm{q})\ph\widehat{\Gamma}_j^{\nu}({\bm{k}+\bm{q},\bm{k}})\tau_{\nu}\right\}\eea
and $\widehat{{\cal G}}(ik_n,\bm{k})$ is the bare matrix Matsubara Green's function
\bea\widehat{{\cal
G}}(ik_n\up,\bm{k})&=&
\frac{ik_n\up-\mu_B\up s_3\up-m_z\up(\bm{k}
)-\mu+\widehat{\bm{g}}({\bm{k}})\cdot\bm{\tau}}{(ik_n\up-\mu_B\up s_3\up-m_z\up(\bm{k})-\mu)^2-E^2({\bm{k}})}
\eea

\noi and the vertex functions $\widehat{\Gamma}_i^{\mu}(\bm{k},\bm{k}+\bm{q})$ are defined as the coefficients of the fermion-gauge field coupling written in the form
$\sum_{\bm{q}}J_{i}^c(\bm{q})A^{i}(-\bm{q})=\sum_{\bm{q},\bm{k}}\Psi_{\bm{k}+\bm{q}}^{\dag}\widehat{\Gamma}_{i}^{\mu}(\bm{k}+\bm{q},\bm{k})\tau_{\mu}\Psi_{\bm{k}}\up
A^{i}(-\bm{q})$. Consequently they are known if we find the fermionic currents. We start by the equation of continuity of the electric charge $Q_{el}\up=\int
d\bm{r}\rho(\bm{r})$ in momentum space $\dot{\rho}(\bm{q})-i\bm{q}\cdot\bm{J}^c(\bm{q})=0$, where we have introduced the electric charge density
$\rho(\bm{q})=-e\sum_{\bm{k}}\Psi^{\dag}_{\bm{k+q}}\Psi_{\bm{k}}\up$. The corresponding electric current can be derived from the equation of continuity as the limit
\bea\bm{J}^c(\bm{q})=-i\lim_{\bm{q}\rightarrow 0}\bm{\nabla_{q}}\ph\dot{\rho}(\bm{q})=\lim_{\bm{q}\rightarrow
0}\bm{\nabla_{q}}\left[\ph{\cal H},\rho(\bm{q})\ph\right]=
-e\sum_{\bm{k}}\Psi^{\dag}_{\bm{k+q}}\widehat{\bm{V}}(\bm{k})\Psi_{\bm{k}}\up\eea

\noi with $\widehat{\bm{V}}(\bm{k})=\bm{\nabla_k}\up\widehat{{\cal H}}(\bm{k})$ the velocity defined in spinor space. Consequently
$\widehat{\Gamma}_i^{0}(\bm{k}+\bm{q},\bm{k})=0$ and $\widehat{\bm{\Gamma}}_i\up(\bm{k}+\bm{q},\bm{k})=\frac{\partial\widehat{\bm{g}}(\bm{k})}{\partial k_i\up}$. 
Straightforward calculation of the polarization tensor yields the Hall topological conductivity
\bea \sigma_{xy}(\omega)=\frac{e^2}{\hbar v}\sum_{\bm{k},s,\nu}\frac{4E^2(\bm{k})\ph n_F\up\left[E_{s,\nu}^{\cal
B}(\bm{k})\right]\Omega_{s,\nu}^z(\bm{k})}{\left[\hbar\omega+i\eta-2E(\bm{k})\right]\left[\hbar\omega+i\eta+2E(\bm{k})\right]}\,.\eea

\noi The imaginary part of the conductivity for $\omega>0$ is given as
\bea \sigma_{xy}^{\Im}(\omega)=-\frac{\pi e^2\hbar\omega}{2\hbar v}\sum_{\bm{k},s,\nu}n_F\up\left[E_{s,\nu}^{\cal
B}(\bm{k})\right]\Omega_{s,\nu}^z(\bm{k})\delta[\hbar\omega-2E(\bm{k})]\,,\eea

\noi where due to the delta function we have the simplification $E_{s,\nu}^{\cal B}(\bm{k})=-\mu-[\mu_Bs+m_z(\bm{k})]{\cal B}+\nu\hbar\omega/2$. Typically, $\hbar\omega$ is of
the order of eV, definitely larger than the energy scales considered here. This means that $n_F\left[E_{s,\nu}^{\cal B}(\bm{k})\right]\simeq n_F[\nu\hbar\omega/2]$. For low
fields, the physics may be well described by considering an expansion around the $\bm{k}_0=(\frac{\pi}{2},\frac{\pi}{2})$ point. In this case, $\Omega_{s,\nu}^z(\bm{k})\simeq
2\nu a^2t\Delta_1\Delta_2/E^3(\bm{k})=16\nu a^2t\Delta_1\Delta_2/(\hbar\omega)^3$, $E(\bm{k})=\sqrt{\Delta_1^2+\Delta_2^2(\delta k_x-\delta k_y)^2+(2t)^2(\delta k_x+\delta
k_y)^2}$, with $\delta\bm{k}=\bm{k}-\bm{k}_0$ and $\frac{1}{v}\sum_{\bm{k}\in{\cal B.Z.}}=\int_{\Delta_1}^{E_c} {\rm dE\ph E}/[2\pi(4t\Delta_2)]$.
\bea \sigma_{xy}^{\Im}(\omega)
=\frac{e^2\Delta_1}{\hbar}\frac{1-2n_F\left(\hbar\omega/2\right)}{\hbar\omega}a^2\simeq \frac{e^2}{\hbar}\frac{\Delta_1}{\hbar\omega}a^2\,.\eea

\end{document}